\documentclass{aa}
\usepackage{graphicx,times}

\begin{document}
 \title{RR Lyrae stars in the outer region of the globular cluster
   M3: a shortage of long periods at r $\sim$ 3.5 to 6 arcmin ?}
\titlerunning{M3: RR Lyrae star periods}

 \author{D. J. Butler
      }
 \offprints{D. Butler}
     \institute{Max-Planck Institut f\"ur Astronomie, K\"onigstuhl 17, D-69117 Heidelberg, Germany \\
              \email{butler@mpia.de} \\
          }
  \institute{Max-Planck-Institut f\"ur Astronomie, K\"onigstuhl 17, D-69117 Heidelberg, Germany \\
              \email{butler@mpia.de}
          }

   \date{Received; accepted}
   \abstract{ 
An analysis of the radial distribution of  ab-type RR Lyrae star periods 
 in the outer region of the 
 globular  cluster M3 at r $\ge$ 0.83$^\prime$ has been
 performed. That analysis points towards a real 
   shortage of stars with long periods  in the radial distance 
 range  3.5$^\prime$
   to  6$^\prime$ (or about 7 to 12 core radii). 
 A brief discussion is presented. The origin of the phenomenon 
 remains an open question.
   }
   \maketitle
%

\section{Introduction}
 Accurate distances to resolved stars in the present-day universe 
 are of  basic
 importance for the  observational 
 study of the stellar content and formation of 
 individual galaxies over time (e.g. Dolphin et al. 2002; Mackey \&
 Gilmore 2003; Gratton et al. 2003). 
  By providing a means to evaluate the
  zero point of the distance scale, RR Lyrae stars,
  in addition
 to classical Cepheid stars,  provide a powerful constraint on the physics of the 
 formation of the Galaxy and Local Group of
  galaxies. For example, for reliable information on 
 stellar masses and their epochs of formation, 
   accurate distances are crucial (e.g. Dolphin et al. 2002).
 There is however a great need to remove 
 systematic errors both in empirical 
 and theoretical
 studies of RR Lyrae stars (e.g. see Carretta et al. 2000; 
 Caputo et al. 2000 and references therein). 
 As their periods  are  more accurately known than other
  parameters (e.g. magnitudes, colours  and amplitudes), 
   the distributions of RR Lyrae star periods in different systems
 should provide a cleaner test of the theoretical models of RR Lyrae stars.
 The purpose of the present paper is to use the database of RR Lyrae star
  periods for M3 from Corwin \& Carney (2001; referred to hereafter as CC01) 
 and to explore the distribution of 
 periods in the outer region of that star cluster.

There have been  past  
 studies of the periods of RR Lyrae stars on different spatial scales.
 For example, on Galactic scales the mean periods of such stars 
 vary systematically with Galactic
 latitude (e.g. Fig.9a in Cseresnjes 2001).
 Differences in period distributions (e.g., mean period and width)
  are also known to occur in Galactic sub-systems and satallite galaxies.
  For example, while period distributions 
 are quite similar in some systems (e.g., Sgr and the Large Magellanic Cloud)
 they are significantly different in others: e.g., 
 compare  Sgr with Carina, Leo II or M15 for instance (Cseresnjes 2001).
  Do  differences occur within any galactic globular clusters ?
  This will be examined in the present paper using
   the  radial distribution of RR Lyrae stars in the outer region of 
  M3, a star cluster  with one of the largest RR Lyrae star populations 
  in the   Galaxy.

The layout of this short paper is as follows. In
  Sec.~\ref{sample1}, the  period data and its radial
  distribution 
  is presented and analyzed. That is followed by a summary of the  result
  and a basic discussion  in Sec.~\ref{discuss}.  
\section{The adopted sample and radial distributions}\label{sample1}
The present study has  made use of  RR Lyrae  star  data 
  for M3 from  CC01.  
  Only ab-type stars at r $\ge$ 0.83$^{\prime}$ have been considered
in order to have an essentially complete sample of 
  periods:
 based on an inspection of the data, completeness 
  at  0.83$^{\prime}$ $\le$ r $\le$  1.7$^{\prime}$
 is about 80\%  and is 100\% 
 at larger radii. Periods
  are precise to  much 
 better than 0.1\%, and the spatial coordinates of isolated stars are
  precise to typically 0.15$^{\prime\prime}$ (Bakos et al. 2000). 

\begin{figure}
\centering
{\includegraphics[angle=0,height=8cm,width=8cm]{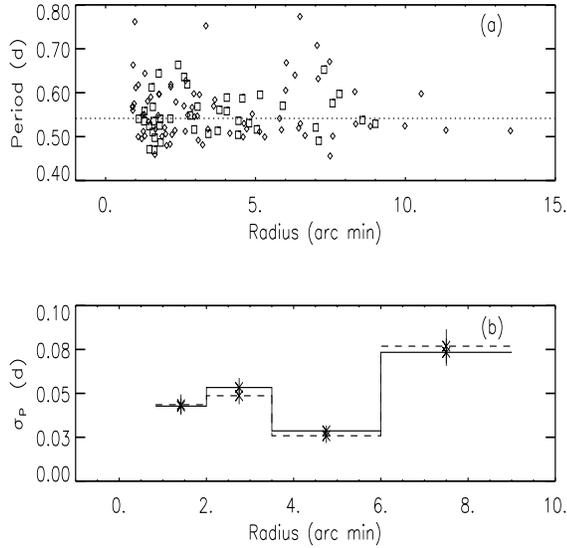}}
\caption{ (a) Radial distribution of  the periods of Blazhko 
 stars (squares), and non-Blazhko ab-type stars (diamonds) at r $\ge$
 0.83$^{\prime}$.   The short-dash line marks the median period of the
 ensemble data. 
 (b) Standard deviation  of periods in annular bins as a function of  
 radial position for  Blazhko plus non-Blazhko stars (solid)  
 and  non-Blazhko stars (dashed). 
 See the text in 
 Sec.~\ref{sample1} for further information. }\label{fig_rad_sig}
\end{figure}

 Fig.~\ref{fig_rad_sig} (a) and (b) plot the 
 radial distribution of periods and their dispersion in four 
 annular bins respectively. Each annular region and 
 its star count   is listed in
 Table~\ref{tableRRLy}. In order to be insensitive to outliers in the data,
  a so-called bootstrap technique has been used to 
    calculate the dispersion in each radial bin and the associated 
  uncertainty:
  80\% of the stars in each radial bin were randomly  selected, and the  dispersion  was recalculated.
 For each radial bin, this procedure was  repeated  several hundred 
  times;  the mean of the resultant set of values 
  was taken as the period dispersion 
   while the standard deviation   was taken
 as the (random) dispersion uncertainty. 
        To determine how much the uncertainty estimates depend on 
  the percentage of stars selected, 
  the percentage selected was varied from 70\% to 90\%: a similar
  mean dispersion value was obtained and  only a slight change in 
 uncertainty was found.

\begin{figure}
\centering
\scalebox{0.5}{\includegraphics[angle=0]{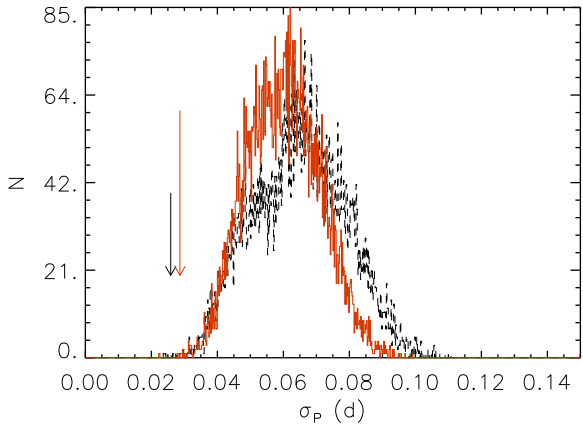}}
\caption{Frequency of occurrence of 
    $\sigma_{\rm P}$  at r = 3.5$^{\prime}$ to 6$^{\prime}$
  for non-Blazhko ab-type stars (dashed) as well
  as Blazhko plus non-Blazhko ab-type
  stars (solid). 
 Arrows indicate the measured dispersion in the same radial range
 for both the former case  (long) and the latter case 
 (short).}\label{fig_prob_RRLy_A_VB} 
\end{figure}

 For an assessment of  the probability that the   measured dispersion 
 at 3.5$^{\prime}$ to 6$^{\prime}$
could have occurred by chance\footnote{ It is noted that a Kolmogorov-Smirnov (K-S) test, which 
 is sometimes used to test the significance of the difference between two
 distributions is not adequate for the  data, unlike the adopted test: 
 the sensitivity
 of the K-S test is best when the determined probability statistic
 does not depend on the tails of the distributions used but rather the 
 central region where number statistics is higher.},
 the following method has been applied:
 if a radial bin contained N stars, then 
  N stars were picked randomly from 
 the whole sample, and the standard deviation was calculated. This 
 was repeated a significant number of times (10,000)
  for each radial bin, thus providing
 a separate histogram of occurrences for each bin.
   The probability of occurrence
   of the measured dispersion values
   is less than 0.01\% for either the set of non-Blazhko\footnote{The Blazhko effect is common in ab-type RR Lyrae stars;  rare in c-type RR Lyrae
  stars;  and is characterized by cycle-to-cycle 
 variations in luminosity and radial velocity curves (i.e. 
 influencing amplitude).}  
 ab-type 
 star periods  or the
 ensemble data. The probability of a value less 
 than or equal to the 
 measured values is 0.03\% and 0.07\% respectively. 
 Accordingly,  this points towards a  shortage of long period
 stars  at  3.5$^{\prime}$ to  6$^{\prime}$ in M3, regardless of
 whether Blazhko stars are included in the analysis or not.
 Histograms for the radial bin at  3.5$^{\prime}$ to  6$^{\prime}$ 
 are shown in Fig.~\ref{fig_prob_RRLy_A_VB}. 
  Vertical arrows indicate the measured dispersion associated with 
  each histogram. 

 For a check of the 
 effect of bin size and bin location,  
 bin sizes at r $\ge$0.83$^{\prime}$ have been scaled 
  simultaneously by  up to 5\%.
 The result is a  small change in dispersion values and 
 negligible change in error bar sizes -- importantly, the
 probability that the radial minimum in dispersion values
  occurred by chance  is still low, being less than 0.1\%.
  As another check, the same data was
 binned at intervals of 1$^{\prime}$ and interestingly, even with reduced
 number statistics, a significant radial minimum occurred in the bin at about
  5$^{\prime}$ to 6$^{\prime}$.

\section{Result and discussion}\label{discuss}
 In the previous section it was shown  clearly 
 that there is 
 evidence for a possible shortage of long period 
 stars of the type ab in the outer region of M3. 
 It  relies on the 
  high precision and accuracy of periods even in the innermost regions,
  unlike  mean colours and amplitudes that may be affected by systematic
 photometric uncertainties due to crowding. 
 It is stated here 
 as only a possible shortage because although random errors are assessed
 well  by
 the boot-strap technique applied earlier, it is not possible
 to rule out the possibility
 of a  systematic error  arising from the finite size of the 
 RR Lyrae star population. 
 With this a caveat, it  is  concluded that 
  the general trend for the dispersion of period values
 is a drop from  the  inner region sampled (0.83$^{\prime}$ to 
   3.5$^{\prime}$)  to  
 smaller values in an intermediate radial 
 region (3.5$^{\prime}$ to  6 $^{\prime}$), 
 and an increase again toward the 
 outer region sampled (beyond about 6$^{\prime}$). 
Beyond a radial distance of about 9$^{\prime}$, there is little
 information due to a decreasing surface density of stars.

Due to the potential importance of the result in 
the context of single star evolution  in star clusters, 
  an
  important parameter worth speculating about briefly is chemical
 composition. 
  Both observationally (e.g. Sandage 1982) and
 theoretically (e.g. Bono et al. 2001) it is known that there is a
 correlation between period, heavy element abundance and
  K-band magnitude, i.e. 
 the PKZ relation\footnote{The  apparent shortage of long periods
 at r = 3.5-6$^{\prime}$ seems 
 to be reflected in the  B- and
 V-band  amplitude data from CC01. It is  expected because 
 of a known correlation between optical amplitudes and periods in M3
 (e.g., Roberts \& Sandage 1955, Fig. 6; CC01, Fig. 9). Herein  an amplitude
  analysis  is excluded  because of likely systematic effects caused by 
  light curve scatter and/or stellar crowding.} 
 Thus, {\it if} taken at face value, the apparent shortage of 
 long periods should be  reflected in K-band magnitudes and/or Z.
  For example, applying the tight  PKZ relation  (e.g. Eqn. (3) of
  Bono et al. (2001)),   one finds that for 
  M$_{\rm K}$ = -0.3\,mag at a mid-way period of 0.5\,d, 
  a period increase of 0.05\,d would correspond to $\Delta$log\,Z $\sim$
 0.6\,dex. The current census of [Fe/H] values is however
 too incomplete to warrant an examination of its radial distribution
  at the present time. 
 
  Another issue worth mentioning 
 is the spatial distribution of RR Lyrae
 stars in M3.
 A Kolmogorov-Smirnov test of the cumulative radial
 distribution of the RR Lyrae stars 
  and  red giant branch (RGB)  stars, using RGB data from Ferraro 
  et al. (1997; and references therein), indicates 
 a probability of 82\%
  that they are from the same parent populations (proof of the
 null-hypothesis). That suggests that the dynamical
 histories of the RR Lyrae star
 and  RGB star populations in M3 are  similar
  in a statistical sense over the range of radial distances
 examined in the present paper. 
 Consequently, if there 
 is some form of selection mechanism that affects RR Lyrae stars,
 and if that mechanism is linked to the dynamics of cluster evolution, 
 its effect on RGB stars, if any, may vary with radial distance in the cluster.

Lastly, although there is no supporting empirical evidence,
 it is of academic interest to sketch very briefly one intriguing
 potential selection mechanism; that is the issue of
  gas giant planets or brown dwarfs that may be interacting with 
 stars evolving along the RGB (Soker 1998). 
 It is believed that  the mass loss rate  of RGB stars would be
  increased by the release of energy and angular momentum 
 on entering the envelope of the RGB star. 
 In the reported scenario, increasing mass loss
 would cause a star to become bluer as it starts
 towards the HB  (Soker 1998).  Speculatively, one could imagine that this
 might preferentially select short RR Lyrae star periods, and that  
 the efficiency of such a mechanism might well vary
 significantly with radial distance in a star cluster.

 \begin{table}
\begin{tabular}{llll}
            \hline
 \rm Annulus         & \rm  N$_{\rm RRab}$$^{\rm a}$   & \rm N$_{\rm non-Blazhko}$
 &  \\
     &  & &  \\
  (1)  &   (2)          &   (3)  \\
 \hline
 0.83$^{\prime}$-2.00$^{\prime}$  & 40 &  25  \\
 2.00$^{\prime}$-3.50$^{\prime}$  & 29 &  21\\
 3.50$^{\prime}$-6.00$^{\prime}$  & 22 &  11    \\
 6.00$^{\prime}$-9.00$^{\prime}$  & 21 & 15  \\
 \hline
\end{tabular}
\noindent 
\\
$^{\rm a}$   Non-Blazhko and Blazhko  ab-type RRLyrae stars
     \caption[]{Numbers of RR Lyrae  
       stars  in different annular regions
  in M3. Columns (2) to (3) have been produced using 
  RR Lyrae star classifications from CC01/Bakos et al. (2000).
  See text in Sec.~\ref{sample1} for further information.}\label{tableRRLy}
\end{table}



\begin{acknowledgements}
 An anonymous referee is thanked  for pertinent comments that helped the author
 to improve  the quality of this paper.  D. J. 
 Butler acknowledges the support of the 
 European research and training network on
 `Adaptive Optics for Extremely Large Telescopes' under contract
  HPRN-CT-2000-00147. 
\end{acknowledgements}

\end{document}